\documentclass{ifacconf}

\usepackage{graphicx}      
\usepackage{natbib}        

\usepackage{geometry}
\usepackage{courier}  
\usepackage{empheq}
\usepackage{amsmath}
\usepackage{amssymb}
\usepackage{mathrsfs}
\usepackage{mathtools}
\usepackage{array}
\usepackage{multirow}
\usepackage{amscd}
\usepackage{xifthen} 
\usepackage{xfrac}
\usepackage{shadow}
\usepackage[inline]{enumitem}
\usepackage{subfigure}
\usepackage{euscript}

\usepackage{matlab-prettifier}

\usepackage{tikz}
\usetikzlibrary{shapes,arrows,shadows.blur,positioning,backgrounds,fit,patterns,calc}  
\usepackage{pgfplots}
\pgfplotsset{compat=newest} 
\pgfplotsset{plot coordinates/math parser=false}

\usepackage{pifont}
\definecolor{iavO}{RGB}{230,150, 44}		
\definecolor{iavH}{RGB}{162,0,103}			
\definecolor{iavB5}{RGB}{ 12, 56,104}		
\definecolor{iavB1}{RGB}{ 93,174,219}		

\newcommand{\lpvcore}{\textsc{LPVcore}}

\newcommand{\Hinf}{\mathcal{H}_{\infty}}    

\DeclareMathOperator*{\argmin}{\arg\!\min~}

\newcommand{\NX}{{n_\mathrm{x}}}
\newcommand{\NXp}[1][]{%
\ifthenelse{\isempty{#1}}{{n_\mathrm{x}}}{{n_\mathrm{x}^{#1}}}%
}

\newcommand{\NY}[1][]{%
\ifthenelse{\isempty{#1}}{{n_\mathrm{y}}}{{n_\mathrm{y}^{#1}}}%
}
\newcommand{\NU}{{n_\mathrm{u}}}
\newcommand{\NP}{{n_\mathrm{p}}}

\newcommand{\NPSI}{{n_{\psi}}}			

\newcommand{\NA}{{n_\mathrm{a}}}
\newcommand{\NB}{{n_\mathrm{b}}}

\newcommand{\NMUTp}[1][]{%
\ifthenelse{\isempty{#1}}{{n_{\tilde{\mathrm{\mu}}}}}{{n_{\tilde{\mathrm{\mu}}}^{#1}}}%
}

\newcommand{\sU}{\mathbb{U}}
\newcommand{\sP}{\mathbb{P}}
\newcommand{\sR}{\mathbb{R}}
\newcommand{\sX}{\mathbb{X}}
\newcommand{\sY}{\mathbb{Y}}

\newcommand{\sT}{\mathbb{T}}

\newcommand{\ddt}{\tfrac{\mathrm{d}}{\mathrm{d}t}}
\newcommand{\ddtp}[1]{\tfrac{\mathrm{d}^{#1}}{\mathrm{d}t^{#1}}}

\newcommand{\Ru}{\mathbb{R}^{\NU}}
\newcommand{\Rp}{\mathbb{R}^{\NP}}
\newcommand{\Rx}{\mathbb{R}^{\NX}}
\newcommand{\Ry}{\mathbb{R}^{\NY}}

\newcommand{\Afnc}{{A}}
\newcommand{\Bfnc}{{B}}
\newcommand{\Cfnc}{{C}}
\newcommand{\Dfnc}{{D}}

\newcommand{\Kfnc}{{K}}

\newcommand{\Dat}{\EuScript{D}_N}

\newcommand{\predDat}[4]{%
\ifthenelse{\equal{#4}{}}{%
\bar{\mathrm{#1}}_{#2}^{{\scriptscriptstyle(\mathrm{#3})}}%
}{%
\bar{\mathrm{#1}}_{#2}^{#4,{\scriptscriptstyle(\mathrm{#3})}}%
}%
}

\DeclareMathOperator{\sinc}{\mathrm{sinc}}

\makeatletter
\newcommand*\bigcdot{\mathpalette\bigcdot@{.5}}
\newcommand*\bigcdot@[2]{\mathbin{\vcenter{\hbox{\scalebox{#2}{$\m@th#1\bullet$}}}}}
\makeatother

\makeatletter
\newcommand{\pushright}[1]{\ifmeasuring@#1\else\omit\hfill$\displaystyle#1$\fi\ignorespaces}
\makeatother   

\newcommand{\norm}[1]{\left\lVert#1\right\rVert}

\DeclareMathAlphabet{\mathpzc}{OT1}{pzc}{m}{it}
\DeclareMathAlphabet{\mathbscr}{U}{BOONDOX-cal}{b}{n}
\DeclareMathAlphabet{\mathdutchcal}{U}{dutchcal}{m}{n}
\DeclareMathAlphabet{\mathbdutchcal}{U}{dutchcal}{b}{n}
\DeclareMathAlphabet{\eucalit}{U}{euf}{m}{n}

\lstMakeShortInline"

\lstdefinestyle{inline}{%
   style              = MATLAB-editor,
  basicstyle         = \ttfamily, 
  escapechar         = ",
  mlshowsectionrules = true,
  breaklines=false,
}

\begin{document}
\begin{frontmatter}

\title{LPVcore: MATLAB Toolbox for LPV Modelling, Identification and Control}

\author[Drebble]{Pascal den Boef}
\author[TNO]{Pepijn B. Cox}
\author[CS,SZTAKI]{Roland T\'{o}th} 

\address[Drebble]{Drebble, Horsten 1, 5612AX, Eindhoven, The Netherlands}
\address[CS]{Control Systems Group, Eindhoven University of Technology, P.O. Box 513, 5600 MB Eindhoven, The Netherlands}
\address[SZTAKI]{Systems and Control Laboratory, Institute for Computer Science and Control, Kende u. 13-17, H-1111 Budapest, Hungary.}
\address[TNO]{Radar Technology, TNO, P.O. Box 96864, 2509 JG The Hague, The Netherlands}
\thanks[footnoteinfo]{This work has received funding from the European Research Council (ERC) under the European Union’s Horizon 2020 research and innovation programme (grant agreement nr. 714663).}
\begin{abstract} 
This paper describes the \lpvcore{} software package for MATLAB developed to model, simulate, estimate and control systems via \emph{linear parameter-varying} (LPV) \emph{input-output} (IO), \emph{state-space} (SS) and \emph{linear fractional} (LFR) representations. In the \lpvcore{} toolbox, basis affine parameter-varying matrix functions are implemented to enable users to represent LPV systems in a global setting, i.e., for time-varying scheduling trajectories. This is a key difference compared to other software suites that use a grid or only LFR-based representations. The paper contains an overview of functions in the toolbox to simulate and identify IO, SS and LFR representations. Based on various prediction-error minimization methods, a comprehensive example is given on the identification of a DC motor with an unbalanced disc, demonstrating the capabilities of the toolbox. The software and examples are available on \lstinline|www.lpvcore.net|. \vspace{-3mm}
\end{abstract}


\end{frontmatter}

\section{Introduction}
\vspace{-3mm}
In recent years, the \emph{linear parameter-varying} (LPV) modelling paradigm has received considerable attention from the identification and control community \cite[e.g., see][]{Toth2010a,Santos2011,Mohammadpour2012,Sename2013,Briat2015,Hoffmann2015}. The LPV framework has become popular as it can represent non-linear and time-varying behavior encountered in real-world systems while, at the same time, the linearity in the model can be exploited to have convex optimization for synthesizing observers and controllers including performance guarantees. An LPV representation describes a linear relation between the inputs and the outputs of the model, but this linear relation is a function of a measurable, time-varying signal, the so-called \emph{scheduling signal} $p$. These variations w.r.t. the scheduling signal enable the representation of non-linear and non-stationary behaviour. Conceptually, LPV models of the underlying system can be obtained by
\begin{enumerate*}[label=\arabic*)]
	\item interpolating various local linear time-invariant representations around fixed operating points, i.e., with constant $p$, often referred to as the \textit{local} or grid based model; or by
	\item formulating one overall representation of the system via the so-called embedding principle, i.e., the \textit{global} form.
\end{enumerate*}
In general, global models capture the dynamics of the system w.r.t. the scheduling signal and, for local models, the scheduling signal is used to interpolate (schedule) between the local linear time-invariant models.

The LPV analysis, modelling, identification, and control literature has become mature in recent years. To accelerate the applicability of LPV methodologies to applications, various computational tools have become available, e.g., LPVTools \citep{Hjartarson2015}, Control Systems and System Identification Toolbox in MATLAB \citep{MATLAB2014}, \emph{Predictor-Based Subspace IDentification} (PBSID) Toolbox \citep{Wingerden2009a}, and \emph{LPV Input/Output Systems Identification Toolbox} (LPVIOID) \citep{Rabeei2015}. The analysis and control methods in the LPVTools and Control Systems toolboxes are mainly focused on a local grid based formulation (Jacobian-linearizations) or a grid based linear fractional representation. The Control Systems Toolbox does include a limited set of tools based on a global form. The PBSID and LPVIOID toolboxes focus on specific algorithms for identification of global LPV models in either SS or IO form, respectively, but with limited options to further utilize these models and with only support for affine dependency on the scheduling signal.

The concept of the initial release of the \lpvcore{} toolbox is to facilitate modelling, simulation, and identification of LPV representations in a global form in MATLAB. Modelling and simulating systems can be performed based on a \emph{continuous-time} (CT) or \emph{discrete-time} (DT) LPV representation. To include the parameter-varying nature, the real and complex number based matrix representations in MATLAB are extended to facilitate parameter-variations. These parameter-varying matrix functions form the core of the toolbox to facilitate algebraic operations. In addition, the \lpvcore{} toolbox includes identification methods to identify discrete-time LPV IO, SS and LFR models. The initial release of the toolbox does not cover the comprehensive LPV literature. However, it provides a solid basis to incorporate LPV analysis and control methods, implement realization tools in a computationally efficient manner, and extend the set of identification methods in the future.

The intention of this paper is to provide an overview of the methods in the initial release of the \lpvcore{} toolbox. In Section~\ref{sec:LPVrep}, the handling of parameter-varying matrix functions and the various LPV representation forms are introduced. The LPV-IO, LPV-SS and LPV-LFR identification methods are described in Section~\ref{sec:ID}. Section~\ref{sec:sim_study} presents an example in which \lpvcore{} is used to estimate a model of a DC motor with unbalanced disc. In Section~\ref{sec:conclusion}, the conclusions are given.
\vspace{-3mm}
\section{LPV representations} \label{sec:LPVrep}
\vspace{-3mm}
In this section, the LPV representations are described to model the system at hand. The key element in formulating the representations in a global form is the parameter-varying matrix object of the \lpvcore{} toolbox, as discussed in Section~\ref{subsec:pmatrix}. With the parameter-varying matrix object, LPV input-output, state-space and linear-fractional representations are constructed, as detailed in Sections~\ref{subsec:LPVIO}, ~\ref{subsec:LPVLFR} and~\ref{subsec:LPVSS}.
\vspace{-3mm}
\subsection{Parameter-varying matrix function} \label{subsec:pmatrix}
\vspace{-3mm}
The LPV representations are defined using \emph{parameter-varying matrix functions}. MATLAB does not have a native implementation of these functions. Therefore, \lpvcore{} introduces the \lstinline|pmatrix| object, defining the following parameter-varying matrix function:
\begin{equation}
(A \diamond p) = A_0 + \sum_{i=1}^{n_{\alpha}} A_i \left(\alpha_{i}\diamond p \right),    \label{eq:parameterPolyA}
\end{equation}
where $p:\sT\rightarrow\sP \subseteq\Rp$ is the scheduling variable, $A_i\in\mathbb{R}^{k\times l}$ are the parameters, and $\alpha_{i}\in\mathscr{R}$ are {scheduling}-dependent coefficient functions.  For the CT case, $\sT=\mathbb{R}$ and, for the DT case, $\sT=\mathbb{Z}$. $\sP$ is a compact {subset} of $\sR^{\NP}$. The set $\mathscr{R}$ is defined as the set of real-analytic functions of the form $f:\sR^{\NP}\times\ldots\times\sR^{\NP}\rightarrow\mathbb{R}$. Furthermore, in the CT case, for a $p\in\sP$, we define the following notation: if $f\in\mathscr{R}$, then $(f\diamond p)=f(p,\ddt p,\ddtp{2} p,\ldots)$ where $\ddt p$ is the time derivative of the scheduling signal. Similarly, in the DT case, for a $p\in\sP$, we define: if $f\in\mathscr{R}$, then $(f\diamond p)=f(p,q^1 p,q^{-1} p,q^{2}p,\ldots)$ where $q$ is the time-shift operator, i.e., $q p_t=p_{t+1}$. Hence, the set $\mathscr{R}$ allows to include various types of non-linearities and time-varying effects into a representation. For example, the affine basis functions $\alpha_{i}\in\mathscr{R}$ can represent polynomials, cosines, exponentials, logarithms, and rational functions in $\ddtp{i} p$ or $p_{t\pm i}$ with $i\geq0$. For the full mathematical treatment of $\mathscr{R}$ see~\citet[Chapter 3]{Toth2010a}. Moreover, for notational simplicity, we denote by $\mathscr{R}^{k\times l}$ the set of all functions of the form~\eqref{eq:parameterPolyA}.

In \lpvcore{}, the \lstinline|timemap| object is introduced to simplify the handling of dynamic dependency on $p$, by introducing the \emph{extended scheduling signal} $\rho$ that includes all the shifted versions (DT) or time derivatives (CT) of $p$ that arise from the use of the $\diamond$ operator in~\eqref{eq:parameterPolyA}. For example, to represent dynamic dependency on both $p_t$ and $p_{t-1}$, the user can generate the extended scheduling signal using \lstinline|timemap([0, -1], 'dt')|. The first argument represents the time shifts or derivatives and the second argument denotes the time domain (\lstinline|'dt'| and \lstinline|'ct'| for DT and CT, respectively).
Then, to model a parameter-varying matrix function in \lpvcore{}, two more ingredients are required:
\begin{enumerate*}
	\item the coefficients $\{A_i\}_{i=0}^{n_\alpha}$, and
	\item the basis functions $\{\alpha_i\}_{i=1}^{n_\alpha}$. 
\end{enumerate*}
To support simple use of parameter-varying matrix functions for casual users, but also support the more advanced needs of researchers, a flexible \lstinline|pmatrix| object is defined, which represents a parameter-varying matrix function based on the three ingredients described above. The first argument of \lstinline|pmatrix| specifies the coefficients, followed by name-value pairs that describe the basis functions and {the} extended scheduling signal. To demonstrate the degree of flexibility in specifying parameter-varying matrix functions, the use of \lstinline|pmatrix| is shown in a few examples:
\begin{exmp}
Consider the following affine parameter-varying matrix function:
\begin{equation}
	\left( A \diamond p \right)_t = A_0 + A_1 p_t + A_2 p_{t-1}.
	\label{eqn:ex1_A}
\end{equation}
The detailed way to represent this function in \lpvcore{} is to first construct a \lstinline|timemap| object to extend the scheduling signal with the time-shifted version, and then use this object to create a \lstinline|pmatrix|, i.e.:
\begin{lstlisting}
rho = timemap([0, -1], 'dt')
A = pmatrix(cat(3, A0, A1, A2), ...
  'BasisType', 'affine', ...
  'BasisParametrization', {0, 1, 2}, ...
  'SchedulingTimeMap', rho)
\end{lstlisting}
Note that the parameter \lstinline|'BasisParametrization'| is a cell array of indices of the extended scheduling signal, with 0 representing the constant factor 1. Indices 1 and 2 refer to $p_t$ and $p_{t-1}$, respectively. As an alternative to \lstinline|pmatrix|, one can simply use \lstinline|preal| and assemble (\ref{eqn:ex1_A}) in one line by algebraic operations (see Example \ref{exm:exm1_continued}).
\end{exmp}
\begin{exmp}
Consider the following affine parameter-varying matrix function with two scheduling signals:
\begin{equation}
	\left( A \diamond (p, q) \right)_t = p_{t-1} + 2 q_{t-1} + 3 p_t + 4 q_t 
\end{equation}
Additional scheduling signals can be specified in the \lstinline|'Name'| option of \lstinline|timemap|. Furthermore, affine basis type is the default and need not be specified when creating the \lstinline|pmatrix|:
\begin{lstlisting}
rho = timemap([-1, 0], 'dt', ...
  'Name', {'p', 'q'})
A = pmatrix(cat(3, 1, 2, 3, 4), ...
  'BasisParametrization', ...
    {1, 2, 3, 4}, ...
  'SchedulingTimeMap', rho)
\end{lstlisting}
\end{exmp}
\begin{exmp}
Consider the following polynomial parameter-varying matrix function:
\begin{equation}
	\left( A \diamond p \right)_t = p_t^{2} + 2 p_{t} p_{t-2} + p_{t-2}^2.
\end{equation}
Setting \lstinline|'BasisType'| to \lstinline|'poly'| specifies polynomial dependence, for which the associated parametrization is a vector with the degree of each term of the extended scheduling signal, i.e.:
\begin{lstlisting}
rho = timemap([-2, 0], 'dt')
A = pmatrix(cat(3, 1, 2, 1), ...
  'BasisType', 'poly', ...
  'BasisParametrization', ...
    {[2, 0], [1, 1], [0, 2]}, ...
  'SchedulingTimeMap', rho)
\end{lstlisting}
\end{exmp}
\vspace{-3mm}
Custom basis functions are supported, too.
In \lpvcore{}, many key operations available in MATLAB for constant matrices, such as taking the sum or product, are extended for \lstinline|pmatrix| objects, providing an intuitive manner to manipulate these objects. The \lstinline|pmatrix| object automatically merges different scheduling signals when these operations are used (see Table \ref{tbl:pmatrix_ops} for an overview).

\begin{table}
	\centering
	\caption{Operators for \lstinline|pmatrix|.}
	\begin{tabular}{|l|l|}
	\hline
		\textbf{Function} & \textbf{Description} \\ \hline
		\lstinline|P1 + P2| & Addition \\ \hline
		\lstinline|P1 - P2| & Subtraction \\ \hline
		\lstinline|P1 .* P2| & Element-wise multiplication \\ \hline
		\lstinline|P1 * P2| & Matrix multiplication \\ \hline
		\lstinline|[P1, P2]| & Horizontal concatenation \\ \hline
		\lstinline|[P1; P2]| & Vertical concatenation \\ \hline
		\lstinline|P^2| & Matrix power \\ \hline
		\lstinline|P.^2| & Element-wise power \\ \hline
		\lstinline|P(:) | & Vectorization \\ \hline
		\lstinline|p = P(i, j)| & Subscripted reference \\ \hline
		\lstinline|P(i, j) = p| & Subscripted assignment \\ \hline
		\lstinline|kron(P1, P2)| & Kronecker product \\ \hline
		\lstinline|P'| & Conjugate transposition \\ \hline
		\lstinline|P.'| & Transposition \\ \hline
		\lstinline|diag([p1, p2])| & Create diagonal matrix \\ \hline
		\lstinline|diag(P)| & Get diagonal elements \\ \hline
		\lstinline|sum(P)| & Sum along columns \\ \hline
		\lstinline|pshift(P, k)| & Shift k time-steps (DT only) \\ \hline
		\lstinline|pdiff(P, k)| & Differentiate k times (CT only) \\ \hline
	\end{tabular}
	\label{tbl:pmatrix_ops}
\end{table}

{For convenience, the function \lstinline|preal| is also available to generate a scalar coefficient}. Combined with \lstinline|pdiff| and \lstinline|pshift| to differentiate or shift the scheduling in CT or DT respectively, it can be used as an alternative syntax for constructing parameter-varying matrix functions, e.g.:
\begin{exmp}
The parameter-varying matrix function (\ref{eqn:ex1_A}) can be constructed by a combination of \lstinline|preal| and algebraic operations in {the} following way:
\begin{lstlisting}
p = preal('p', 'dt')
A = A0 + A1 * p + A2 * pshift(p, -1)
\end{lstlisting}
\label{exm:exm1_continued}
\end{exmp}
\vspace{-3mm}
As highlighted by the above examples, an important strength of \lpvcore{} is the implementation of the parameter-varying matrix functions and, as shown in the next sections, the ability to model, analyze, identify, and control LPV representations in a global form. Local representational forms can be obtained via conversion functions, which will be introduced for compatibility with LPVTools in the future.
\vspace{-3mm}
\subsection{LPV-IO representation} \label{subsec:LPVIO}
\vspace{-3mm}
With the parameter-varying matrix function defined, various forms of LPV representations of a system at hand can be formulated. A commonly used representation form is the \emph{input-output} (IO) representation, given by the following equation:\vspace{-1mm}
\begin{equation}
(\mathcal{A}(\xi)\diamond p)_t\ y_t=(\mathcal{B}(\xi)\diamond p)_t\ u_t, \label{eq:IOProcess} \vspace{0.5mm}
\end{equation}
where $y:\sT\rightarrow\sY=\Ry$ is the measured output signal and $u:\sT\rightarrow\sU=\Ru$ denotes the input signal. For the CT case, $\xi=\ddt$, i.e., $\xi y_t=\ddt y_t$ denotes the time derivative of the output; and for the DT case, $\xi= q^{-1}$ is the backward time-shift operator. The {scheduling}-dependent polynomials $\Afnc(\xi)$, $\Bfnc(\xi)$ are polynomials in the indeterminate $\xi$ with {scheduling-dependent} coefficients:\vspace{-0mm}
\begin{subequations}  \label{eq:polyProcess}
\begin{align}
\mathcal{A}(\xi)\diamond p&=I_{\NY}+\sum_{i=1}^\NA (A_i\diamond p)\, \xi^{i}, \\[-2mm]
\mathcal{B}(\xi)\diamond p&=\sum_{j=0}^\NB (B_j\diamond p)\, \xi^{j},
\end{align}
\end{subequations}
with finite polynomial orders $\NA\geq0$ and $\NB\geq0$, respectively, and $A_i\in\mathscr{R}^{\NY\times\NY}$ and $B_j\in\mathscr{R}^{\NY\times\NU}$ are  scheduling-dependent matrix coefficient functions that are affine in the basis functions $\{\alpha_{i,l} \}_{i=1,l=1}^{\NA,n_{\alpha}}$, $\{\beta_{j,l} \}_{j=0,l=1}^{\NB,n_{\beta}}$, respectively,  similarly to~\eqref{eq:parameterPolyA}. The LPV-IO representation~\eqref{eq:IOProcess} can be created by using the \lstinline|lpvio| command.
\vspace{-3mm}
\subsection{LPV-SS representation} \label{subsec:LPVSS}
\vspace{-3mm}
The system can also be represented by a first-order differential or difference equation, i.e., state-space representation:
\begin{subequations} \label{eq:SSrepDet}
\begin{alignat}{3}
  \xi x_t &= ({A} \diamond p)_t\,&x_t&+(\Bfnc\diamond p)_t&&u_t, \label{eq:SSrepDetState} \\
  y_t &= (\Cfnc\diamond p)_t&x_t&+(\Dfnc\diamond p)_t&&u_t. \label{eq:SSrepDetOut}
\end{alignat}
\end{subequations}
where $x:\sT\rightarrow\sX=\Rx$ is the so-called state (latent) variable and $\Afnc\in\mathscr{R}^{\NX\times\NX}$, $\Bfnc\in\mathscr{R}^{\NX\times\NU}$, $\Cfnc\in\mathscr{R}^{\NY\times\NX}$, and $\Dfnc\in\mathscr{R}^{\NY\times\NU}$ are matrix coefficient functions. For the CT case, $\xi=\ddt$ and, for the DT case, $\xi= q$ is the forward time-shift operator. The LPV-SS representation (\ref{eq:SSrepDet}) can be created by using the \lstinline|lpvss| command. 
\vspace{-3mm}
\subsection{LPV-LFR representation} \label{subsec:LPVLFR}
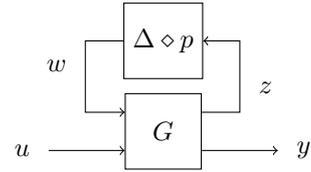
\begin{figure}
	\centering
	\begin{tikzpicture}[scale=1, node distance=1.2cm]
		\node (Delta) [draw, minimum size=1cm] {$\Delta \diamond p$};
		\node (G) [draw, below of=Delta, minimum size=1cm] {$G$};
		
		\draw[->] (Delta.west) -| ++(-5mm,0) node [label=below left:{$w$}] {} |- ($(G.north west)!.5!(G.west)$);
		\draw[->] ($(G.north east)!.5!(G.east)$) -| ++(5mm,0) node [label=above right:{$z$}] {} |- (Delta.east);
		\draw[->]($(G.south west)!.5!(G.west) - (10mm, 0)$) node [label=left:{$u$}] {} -- ($(G.south west)!.5!(G.west)$);
		\draw[->] ($(G.south east)!.5!(G.east)$) -- ($(G.south east)!.5!(G.east) + (10mm, 0)$) node [label=right:{$y$}] {};
	\end{tikzpicture}
	\caption{Schematic representation of an LPV-LFR model.}
	\label{fig:lpvlfr}
\end{figure}

\vspace{-3mm}
An LPV-LFR representation can be used if the dynamics of the system can be decomposed as the linear fractional transform (LFT) of an LTI system $G$ with a parameter-varying matrix function $\Delta$, as shown in Figure \ref{fig:lpvlfr}. The input-output behavior of the system can then be described by the following equations:
\begin{subequations}
\begin{alignat}{4}
  \xi x_t &= \Afnc\,x_t +\Bfnc_\mathrm{w} w_t && + \Bfnc_\mathrm{u} u_t, \label{eq:LFRrepState} \\
  z_t &= \Cfnc_\mathrm{z} x_t +\Dfnc_\mathrm{zw} w_t && + \Dfnc_\mathrm{zu} u_t, \label{eqn:LFRrepZOut} \\
  y_t &= \Cfnc_\mathrm{y} x_t +\Dfnc_\mathrm{yw} w_t && + \Dfnc_\mathrm{yu} u_t, \label{eq:LFRrepYOut} \\
  w_t &= (\Delta\diamond p)_t z_t.
\end{alignat}
\end{subequations}
The LPV-LFR representation can be created by using the \lstinline|lpvlfr| command. The LPV-LFR representation is a generalization of the LPV-SS representation, as can be seen by setting $\Dfnc_\mathrm{zw} = 0$. Therefore, \lpvcore{} internally treats LPV-SS models as LPV-LFR models, and every operation defined on LPV-LFR models, such as multiplication, concatenation and feedback interconnection, can directly be used on LPV-SS models. For a complete list of supported operations, see \lstinline|www.lpvcore.net|.

\vspace{-3mm}
\section{Identification of LPV models} \label{sec:ID}
\vspace{-3mm}
\lpvcore{} includes several system identification methods to estimate DT LPV-IO and LPV-SS models in a global setting, and one method to estimate LPV-LFR models in a local setting. For the global approach, the estimation of the model parameters is performed by minimizing the prediction-error. Hence, in Section~\ref{subsec:PEM}, the \emph{prediction-error minimization} (PEM) setting is briefly introduced. Next, the LPV-IO model set and the corresponding identification methods are discussed in Section~\ref{subsec:IOID}. In Section~\ref{subsec:SSID}, the model set and PEM for LPV-SS models are discussed. For the local approach, to identify LPV-LFR models, the estimation is based on minimizing the worst-case {local} $\Hinf$-error between the model and a given set of LTI models at the measured operating points. This approach is discussed in Section~\ref{subsec:LFRID}.
\vspace{-3mm}
\subsection{The PEM identification setting} \label{subsec:PEM}
\vspace{-3mm}
This section provides a short overview of PEM, for a more detailed description, see e.g. \cite{CHToth2012}. In PEM, model estimates are obtained by minimizing their associated prediction-error with respect to a loss function where the (unknown) model parameters are the optimization variables.  Hence, the concept is to obtain a parameter estimate $\hat\theta_N$ of the parameter $\theta_\mathrm{o}$ describing the data-generating system by minimizing the least-squares criterion, i.e., by
\begin{equation}
\hat{\theta}_N=\argmin_{\theta\in\mathbb{R}^{n_{\theta}}}V(\Dat,\theta), \label{ch6-eq:parameter_Est}
\end{equation}
with 
\begin{equation*}
V(\Dat,\theta)=\frac{1}{N}\sum_{t=1}^{N} \|\varepsilon_{t | \theta}\| _2^2= \frac{1}{N} \| \varepsilon_\theta \| _{\ell_2}^2,
\end{equation*}
where  $\Dat = \{u_t, p_t, y_t\}_{t=1}^N$ is the observed data sequence of the system and the one-step-ahead prediction-error is $\varepsilon_{t | \theta}=y_t-\hat{y}_{t | \theta, t-1}$. In Sections~\ref{subsec:IOID} and~\ref{subsec:SSID}, the one-step-ahead prediction-error $\varepsilon_{t | \theta}$ associated with the IO and SS model sets will be introduced. 

In system identification, the data $\Dat$ is assumed to be corrupted with noise. Hence, it is essential to extend the system representations with a noise process. These extensions are discussed in the next sections.
\vspace{-3mm}
\subsection{LPV-IO model identification}  \label{subsec:IOID}
\vspace{-3mm}
\subsubsection{Model set{:}}
The IO representation of the data-generating system is captured by the parametrized model $\mathcal{M}_\theta$:
\begin{subequations} \label{eq:totalmodel}
\begin{align}
(\mathcal{F}(q^{-1},\theta)\diamond p)_t\ \breve{y}_t &= (\mathcal{B}(q^{-1},\theta)\diamond p)_t\,q^{-\tau_\mathrm{d}} \ u_t, \label{eq:totalmodelp} \\
(\mathcal{D}(q^{-1},\theta)\diamond p)_t\ v_t&= (\mathcal{C}(q^{-1},\theta)\diamond p)_t\ \varepsilon_{t|\theta}, \label{eq:totalnoisem}\\
(\mathcal{A}(q^{-1},\theta)\diamond p)_t\ y_t &=\breve{y}_t+v_t,
\end{align}
\end{subequations}
where $\mathcal{A}(\centerdot),\ldots,\mathcal{D}(\centerdot),\mathcal{F}(\centerdot)$ are polynomials in $q^{-1}$ with {scheduling-}dependent coefficients, which are parametrized in $\theta$ using a-priori {given} basis functions.

The model~\eqref{eq:totalmodel} with $\mathcal{A}\triangleq I_{\NY}$ is also known as the LPV \emph{Box-Jenkins} (BJ) model. By considering $\mathcal{C}\triangleq \mathcal{D} \triangleq \mathcal{F}\triangleq I_{\NY}$ in \eqref{eq:totalmodel}, the LPV version of the so-called \emph{auto regressive with exogenous input} (ARX) model structure is obtained. Similarly, the \emph{auto regressive moving average with exogenous input} (ARMAX) model is found by considering $\mathcal{D}\triangleq \mathcal{F}\triangleq I_{\NY}$, and the \emph{output-error} (OE) model by $\mathcal{A}\triangleq\mathcal{C}\triangleq\mathcal{D}\triangleq I_{\NY}$.

Each of the above model structures is available in \lpvcore{} using the \lstinline|lpvidpoly| object, which serves as an LPV extension to the \lstinline|idpoly| object included in the System Identification Toolbox. A powerful feature of \lstinline|lpvidpoly| is the ability to exploit known sparsity of $\theta$, as demonstrated in the following example.
\begin{exmp}[LPV-ARX model]
Consider an LPV-ARX model of the form $A(p,{\theta}) q^{-1} y + y = B{(\theta)} u$:
\begin{equation}
	A(p,{\theta}) = \begin{pmatrix} \theta_1 & \theta_2 \\ 0 & \theta_3 \end{pmatrix} p, \quad
	B{(\theta)} = \begin{pmatrix} \theta_4 \\ 0 \end{pmatrix}.
\end{equation}
This model structure can be created as follows:
\begin{lstlisting}
p = preal('p', 'dt');
A = [1, 1; 0, 1] * p;
B = [1; 0];
sys = lpvidpoly({eye(2), A}, B);
\end{lstlisting}
\end{exmp}
The object \lstinline|sys| is then used as a \emph{template model structure} in subsequent LPV-IO {identification} methods, which use it to extract the scheduling dependence and, in some methods, the initial parameter values. Zero elements are, by default, excluded from the set of identifiable parameters, providing an intuitive method for semi-grey box identification of LPV-IO representations. This behaviour can be disabled: see documentation at \lstinline|www.lpvcore.net|.
\subsubsection{Identification schemes{:}}
The toolbox contains four time-domain methods for identifying IO model structures~\eqref{eq:totalmodel}: 1) linear regression, 2) pseudo-linear regression, 3) gradient-based search, and 4) instrumental variable method. These methods are based on~\cite{Butcher2008,CHToth2012,Cox2018PHD}. Linear regression is applicable for the ARX model set and the model estimate is obtained by executing \lstinline|lpvarx|. For the \lstinline|lpvarx| method, regularization is included by solving a Tikhonov regression problem, where various weighting matrices can automatically be tuned using generalised cross-validation \citep[Section 6.1.4]{Golub2013} or marginal likelihood optimization \citep{Rasmussen2006}. These options can be selected using the option set \lstinline|lpvarxOptions|.  The pseudo-linear regression algorithm is executed for the ARMAX, OE, and BJ model sets by calling \lstinline|lpvarmax|, \lstinline|lpvoe|, and \lstinline|lpvbj|, respectively. The gradient-based search is applicable for all afore-mentioned model types and is executed by calling \lstinline|lpvpolyest|. The instrumental variable method is applied by the command \lstinline|lpviv|. At the moment, this function is available for model structures with a single input and a single output (SISO), and will be extended to the general case in a later release.
\vspace{-3mm}
\subsection{LPV-SS model identification}  \label{subsec:SSID}
\vspace{-3mm}
\subsubsection{Model set{:}}
The model set for LPV-SS representations supports \emph{innovation noise}, i.e.:
\begin{subequations} \label{eq:SSrep}
\begin{alignat}{3}
  x_{t+1} &\!=\! (\Afnc(\theta)\diamond p)_t&x_t&\!+\!(\Bfnc(\theta)\diamond p)_t&&u_t\!+\!(\Kfnc(\theta)\diamond p)_t e^\mathrm{o}_t, \label{eq:SSrepState} \\
  y_t &\!=\! (\Cfnc(\theta)\diamond p)_t&x_t&\!+\!(\Dfnc(\theta)\diamond p)_t&&u_t\!+\!e^\mathrm{o}_t, \label{eq:SSrepOut}
\end{alignat}
\end{subequations}
where the matrix coefficient functions $\Afnc,\ldots,\Dfnc,\Kfnc$ are affine in the set of a-priori specified basis functions $\left\{\psi_i\diamond p\right\}_{i=1}^\NPSI$ similar to~\eqref{eq:parameterPolyA}. The sequence $e^\mathrm{o}_t$ is the sample path realization of the zero-mean stationary process:
\begin{equation} \label{eq:noisyGB}
\mathbf e^\mathrm{o}_t\sim\mathcal{N}(0,\Xi),
\end{equation}
with $\Xi$ a positive-definite, symmetric real matrix.

\subsubsection{Identification schemes{:}}
To identify the state-space model with parametrization~\eqref{eq:SSrep}, the toolbox includes an iterative, \emph{gradient-based} (GB) optimization technique which obtains an estimate of the innovation model. In a future release, a second iterative optimization technique which is based on the \emph{expectation-maximization} (EM) (\cite{NN:11}) method will be included. The \lstinline|lpvssest| function estimates an LPV-SS model with the GB algorithm. Since this method is based on a non-linear optimization, an initial estimate must be provided by the user. This initial estimate can be obtained by identifying an IO model and transforming it into the state-space form (available in a future release of \lpvcore{}) or {by subspace identification which is the default option} \citep{Toth21aAUT,Wingerden2009a}.
The GB method includes several steps to efficiently estimate the parametric LPV-SS model with innovation noise~\eqref{eq:SSrep}. Among others, the implementation ensures that the non-linear optimization does not wander among parametrizations of SS models with equivalent input-scheduling-output behavior. 
The method is explained in detail in~\citet[Sec. 8.2]{Cox2018PHD}.
\vspace{-2mm}
\subsection{LPV-LFR model identification} \label{subsec:LFRID}
\vspace{-2mm}
The local identification scheme for LPV-LFR representations included in \lpvcore{} is based on \cite{vizer2013linear}. Due to a lack of space, a detailed explanation can be found on \lstinline|www.lpvcore.net| soon.
\vspace{-2mm}
\subsection{Using identified models for control synthesis} \label{ssec:control}
\vspace{-2mm}
\lpvcore{} includes several control synthesis algorithms. LPV-SS models identified using the previously introduced schemes can directly be used for control by synthesizing a stabilizing controller {based on given performance  specifications} in a variety of settings. For a complete overview of the available algorithms, consult the documentation which is included in \lpvcore{}.
\vspace{-2mm}
\section{Simulation study} \label{sec:sim_study}
\vspace{-2mm}
\subsection{Non-linear data-generating system}
\vspace{-2mm}
To illustrate the effectiveness of \lpvcore{}, several of the implemented identification methods are applied to estimate a model of a DC motor with an unbalanced disc. The system consists of a vertically mounted disc with a lumped mass $m$ at distance $l$ from the center, actuated by an input voltage $u$. Due to the gravitational force exerted on the lumped mass, the system exhibits non-linear dynamics, which can be represented in the following differential equation:
\begin{equation}
	\ddot{\theta} = -\frac{1}{\tau} \dot{\theta} + \frac{K_\mathrm{m}}{\tau} u - \frac{m g l}{J} \sin(\theta)
	\label{eqn:unb_disc_nonlinear}
\end{equation}
where $\theta$ is the disc angular position, $\tau$ is the lumped back EMF constant, $K_{m}$ is the motor constant, $J$ is the complete disc inertia and $g$ is the acceleration due to gravity. The goal of this section is to identify the dynamics of (\ref{eqn:unb_disc_nonlinear}) based on simulation data. The {simulated} output is the angular position $\theta$ corrupted with additive white Gaussian noise $\varepsilon$ of a specified  {\emph{signal-to-noise ratio}} SNR. 
\vspace{-3mm}
\subsection{LPV identification}
\begin{figure}[t]
	\centering
	\includegraphics[width=1\linewidth]{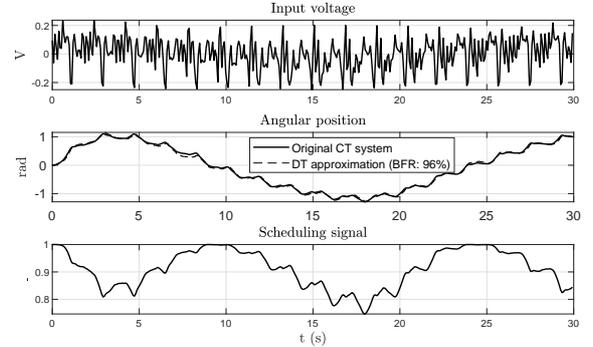}\vspace{-3mm}
	\caption{{Estimation dataset from the unbalanced disc system (\ref{eqn:unb_disc_nonlinear}), compared to the output of  its DT approximation (\ref{eqn:unb_disc_lpvio}).}}
	\label{fig:datasets}
\end{figure}

\vspace{-3mm}
An LPV embedding of the non-linear dynamics in (\ref{eqn:unb_disc_nonlinear}) can be obtained by introducing the scheduling signal $p$:
\begin{equation}
	p := \sinc(\theta) = \frac{\sin(\theta)}{\theta}.
	\label{eqn:unb_disc_p}
\end{equation}
Substitution of (\ref{eqn:unb_disc_p}) into (\ref{eqn:unb_disc_nonlinear}) leads to the following LPV-IO representation of the dynamics of the unbalanced disc:
\begin{equation}
	\ddot{\theta} = -\frac{1}{\tau} \dot{\theta} - \frac{m g l}{J} p \theta + \frac{K_\mathrm{m}}{\tau} u
	\label{eqn:unb_disc_lpv}
\end{equation}
Global identification approaches available in \lpvcore{} operate in DT. Therefore, the LPV system (\ref{eqn:unb_disc_lpv}) is discretized using forward Euler, i.e.:
\begin{equation}
	\dot{\theta}_k := \dot{\theta}(k T_\mathrm{s}) \approx \frac{\theta_{k+1} - \theta_{k}}{T_\mathrm{s}},
	\label{eqn:forward_euler}
\end{equation}
with {$T_\mathrm{s}=0.75$ ms} the sampling time. Applying (\ref{eqn:forward_euler}) to (\ref{eqn:unb_disc_lpv}) yields the following DT approximation:
\begin{equation}
	\theta_{k} + A_1 \theta_{k-1} + 
		\left( A_2 \diamond p \right)_k \theta_{k-2}
	= \frac{K_\mathrm{m} T_\mathrm{s}^2}{\tau} u_{k-2},
	\label{eqn:unb_disc_lpvio}
\end{equation}
with
\begin{equation}
	A_1 = \frac{T_\mathrm{s}}{\tau} - 2, \quad
	(A_2 \diamond p)_k = 1 - \frac{T_\mathrm{s}}{\tau} + \frac{m g l T_\mathrm{s}^2}{J} p_{k-2}.  \\
\end{equation}
Figure \ref{fig:datasets} shows the quality of the approximation in terms of the \emph{Best Fit Rate} (BFR) for a simulation. The BFR, expressed as a percentage, is defined as follows:
\begin{equation}
	\mathrm{BFR} := 100\% \cdot \max \left( 0, 1 - \frac{\norm{y - \hat{y}}_2}{ \norm{ y - \mathrm{mean}(y)}} \right).
\end{equation}
The values of all numerical parameters used in this example are listed in Table \ref{tbl:unb_disc_parameters}.

\begin{table}[t]
	\centering
	\caption{Parameters of the unbalanced disc.}
	\begin{tabular}{|l|l|l|l|}
	\hline
		\textbf{Parameter} & \textbf{Symbol} & \textbf{Value} & \textbf{Unit} \\
	\hline
		Sampling time & $T_\mathrm{s}$ & 75 & ms \\
		Motor constant & $K_\mathrm{m}$ & 15.3145 & - \\
		Disc inertia & $J$ & $2.2 \cdot 10^{-4}$ & $\mathrm{Nm}^2$ \\
		Lumped mass & $m$ & 0.07 & kg \\
		Distance mass to center & $l$ & 0.42 & mm \\
		Gravitational acceleration & $g$ & 9.8 & $\mathrm{m} \cdot \mathrm{s}^{-2}$ \\
		Back EMF constant & $\tau$ & 0.5971 & - \\ 
	\hline
	\end{tabular}
	\label{tbl:unb_disc_parameters}
\end{table}

For the identification procedure, five datasets of 400 samples each are generated: four for estimation with varying levels of additive White Gaussian output noise and one for validation (without noise). These datasets are generated by simulating (\ref{eqn:unb_disc_nonlinear}) for a random-phase multi-sine input $u$ containing 10 frequencies equally spaced in a passband from 0 to 0.75 times the Nyquist frequency. The signal is scaled to have a maximum amplitude of 0.25 V. Zero-order-hold is used to interpolate between the samples. The SNRs for the estimation datasets are 0, 10, 20 and 40 dB. A noise-free example of the estimation dataset and a comparison with the DT LPV-IO approximation (\ref{eqn:unb_disc_lpvio}) can be seen in Figure \ref{fig:datasets}. Four LPV-IO model structures are identified: ARX, ARMAX, OE and BJ. It is assumed that the true model structure (\ref{eqn:unb_disc_lpvio}) is not known exactly and the following more generic {template} structure is used instead (see (\ref{eq:totalmodel})):
\begin{itemize}
	\item $(\mathcal{A}(q^{-1}) \diamond p)_k = 1 + \left( 1 + p_{k-1} \right) q^{-1} + \left( 1 + p_{k-2} \right) q^{-2}$
	\item $(\mathcal{B}(q^{-1}) \diamond p)_k = (\mathcal{F}(q^{-1}) \diamond p)_k = (\mathcal{A}(q^{-1}) \diamond p)_k$
	\item $\mathcal{C}q^{-1}) = \mathcal{D}(q^{-1}) = 1 + q^{-1}$
\end{itemize}
The models are identified using the estimation dataset in the following order:
\begin{enumerate*}
	\item The {LPV-ARX} model is identified by calling \lstinline|lpvarx|.
	\item The {LPV-ARMAX} and {LPV-OE} models are identified by calling \lstinline|lpvpolyest|. The LPV-ARX model is used as initialization.
	\item The {LPV-BJ} model is identified by calling \lstinline|lpvpolyest|. The LPV-OE model is used as initialization.
\end{enumerate*}
For each method, the default options were used, except the number of gradient descent iterations performed by \lstinline|lpvpolyest|: it was set to 400. The quality of the resulting models is evaluated using the validation dataset by calculating the BFR between the output of the non-linear CT system (\ref{eqn:unb_disc_nonlinear}) and the output of the DT model estimate.
\vspace{-3mm}
\subsection{Results}
\begin{figure}
	\centering
	\includegraphics[width=0.5\textwidth]{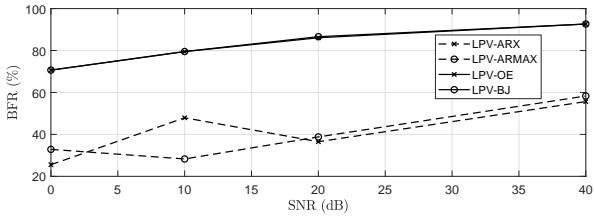}
	\caption{Validation BFR of the model estimates for different output SNRs of the estimation dataset.}
	\label{fig:bfr_snr}
\end{figure}

\begin{figure}
	\centering
	\includegraphics[width=0.5\textwidth]{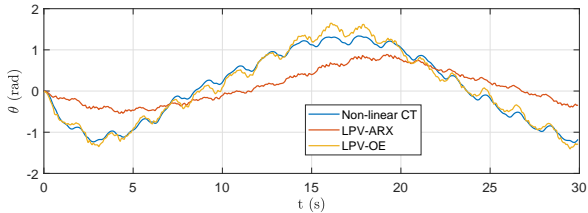}
	\caption{Comparison between the output of the validation dataset and the DT LPV-ARX and LPV-OE models estimated with SNR = 20 dB.}
	\label{fig:validation_snr20dB}
\end{figure}

\vspace{-3mm}
The achieved BFR of the model estimates can be seen in Figure \ref{fig:bfr_snr}. For each SNR, the LPV-OE and LPV-BJ models have a substantially higher BFR than the LPV-ARX and LPV-ARMAX models, which is in line with the expectation since output noise was used. The low BFR of the LPV-ARX and LPV-ARMAX models, even for high SNRs, is likely the result of a high parameter sensitivity due to the proximity of the poles of the frozen dynamics of (\ref{eqn:unb_disc_lpv}) to the unit circle. For example, at $p = 0$ ($\theta = \pm \pi$ rad), the magnitude of the two poles of the resulting LTI dynamics are 1 and 0.92. When $p = 1$ ($\theta = 0$), both poles have a magnitude of 0.96. The output of the LPV-ARX and LPV-OE models estimated with SNR = 20 dB are compared to the original non-linear CT dynamics (\ref{eqn:unb_disc_nonlinear}) in Figure \ref{fig:validation_snr20dB}.

\vspace{-3mm}
\section{Conclusion} \label{sec:conclusion}
\vspace{-3mm}
This paper introduces a software package for MATLAB called the \lpvcore{} toolbox to model, simulate, and estimate linear parameter-varying input-output, state-space and linear fractional representations in a global setting. To support the global setting, the package defines parameter-varying matrix functions, which is the key difference to other software suites that use a grid or linear fractional based representation. This paper provides an overview of the core modelling tools combined with the simulation and identification methods. The software and examples are freely available on \lstinline|www.lpvcore.net|.

The initial toolbox forms a solid basis for additional algorithms. In the near future, algorithms for LPV realization and {a wide range of subspace identification methods will be implemented}.
\vspace{-2mm}
\bibliography{library}

\end{document}